\documentclass{article}

\usepackage{epsf,hyperref}
\usepackage{amssymb,ComplexSystems}
\usepackage{graphicx}%
\usepackage{algorithm}%
\usepackage{algorithmicx}%
\usepackage{algpseudocode}%

\begin{document}

\title{Musical composition and 2D cellular automata based on music intervals%
%
}

\author{\authname{Igor Lugo}\\[2pt] 
\authadd{Centro Regional de Investigaciones Multidisciplinarias (CRIM)}\\\authadd{Universidad Nacional Auton\'{o}ma de M\'{e}xico (UNAM)}\\
\authadd{Av. Universidad s/n, Circuito 2}\\
\authadd{Cuernavaca, Morelos, 62210, M\'{e}xico}\\
\and
\authname{Martha G. Alatriste-Contreras}\\
\authadd{Facultad de Econom\'{i}a}\\ 
\authadd{Universidad Nacional Auton\'{o}ma de M\'{e}xico (UNAM)}\\
\authadd{Circuito Escolar, C.U. Coyoac\'{a}n}\\
\authadd{Ciudad de M\'{e}xico, 04510, M\'{e}xico}
}

%
%

\maketitle

\begin{abstract}
This study is a theoretical approach for exploring the applicability of a 2D cellular automaton based on melodic and harmonic intervals in random arrays of musical notes. The aim of this study was to explore alternatives uses for a cellular automaton in the musical context for better understanding the musical creativity. We used the complex systems and humanities approaches as a framework for capturing the essence of creating music based on rules of music theory. Findings suggested that such rules matter for generating large-scale patterns of organized notes. Therefore, our formulation provides a novel approach for understanding and replicating aspects of the musical creativity.
\end{abstract}


\section{Introduction}\label{intro}
Musical composition can be defined as the process of creating music through a mental activity that generates new ideas and procedures by working on existing ones. Musicians have explored different strategies for conceiving a piece of music, for example following a more traditional approach, musicians can start from a simple melody or chord progression to complex orchestration using music notation or musical instrument performances. On the other hand, nonmusicians can spontaneously invent pieces of music by expressing their musicality \cite{WeissandPeretz2022}. 
However, the musical composition based on simple rules used in computer programs has been slightly altered and explored since the work of Xenakis \cite{Xenakis1992} and Wolfram \cite{Wolfram2002}. Based on cellular automata (CA) for composing music, Xenakis was interested in how simple computational rules applied to music may create complex structures related to harmonic progressions and instrument orchestrations. In the work of Wolfram, he suggested that such simple programs can generate rich and complex behaviors possibly associated with pleasing pieces of music. Then, both formulations create music based on generated complex structures related to particular computational rules. However, the exploration of new ways of using CA based on musical theory has been little analyzed. Therefore, we aim to investigate new applications of CAs by using specific concepts of music theory. In particular, we are interested in harmonizing random arrays of musical notes by using particular rules of melodic and harmonic intervals. This type of strategy for composing may closely resemble the traditional forms of musical composition. The mental process associated with the musical creativity can be captured in a few lines of code related to simple musical rules. Then, this study provides the musical perspective for complementing the current knowledge of musical composition and creativity by incorporating the use of computational models as instruments for producing new ideas in music.

Implementing CAs in the musical composition for understanding and replicating aspects of the human artistic expression is not trivial. However, based on a musical perspective, we can connect the music theory with the computational rules used in the CAs. In particular, following the formulation of the CAs, the state of each cell of some particular array may consist of a note related to the scientific pitch notation (SPN), for example $A4$, which is related to the Western system of music. This note shall be updated according to some particular rules that should use music intervals. Because intervals measures the distance between notes---usually in terms of half-steps and whole-steps---we can use them for setting the neighborhood around a particular note and the state of every neighbor cell. Then, using a set of rules based on intervals, we can update the state of each cell and, then, the whole array simultaneously. Therefore, the proximity of notes in music theory and practice can be translated into the neighborhood of each cell. By this means, we can explore an alternative way of understanding musical creativity by harmonizing random sets of notes in a particular tonal center. Hence, we can reveal 
how the human intuition related to the creativity process in composing music can be translated into a computer code for identifying well-defined procedures that can be executed by a set of music rules. In addition, we aim to show the dynamics from disorder to order phases associated with a particular type of musical system \cite{Berezovsky2019}. 

This study is set out to better understand the use of CAs for assisting musicians---from classical to contemporary approaches---in the musical composition based on a computational formulation. Therefore, our main questions are the following: Can the CAs based on rules of music theory coordinate at large-scale sets of random notes in a particular key? Can the use of transitional rules based on music intervals resemble the traditional approach of music for composing music?
To answer these questions, we aim to use the complex systems approach and the humanities discipline for studying the field of musical composition. In particular, the complex systems provides the framework for understanding the musical composition as a system formed by components and their relationships in which components can be related to the music theory and practice. Then, their interactions can give rise to generating pieces of music. In this vein, we selected the CAs as our best approximation for modeling musical composition based on complex systems because the CAs can incorporate aspects of melody and harmony associated with intervals, as well as other musical attributes such as rhythm and note duration. In addition, the discipline of humanities contributes with its emphasis of understanding the human creativity in the process for creating music. Therefore, we propose this type of CAs that complement the process of creating music by assisting musicians to consider any type of rule related to music intervals. 
We hypothesized that the use of CAs in the musical composition may provide new insights into the human creativity by showing one aspect of the process behind it. In particular, we state that the use of this type of computational models in music can replicate one of the most common strategies by musicians for composing music. Their conscious and unconscious processes for determining a sequence of actions, as they play an instrument and compose music, are closely related to the use of rules in a particular CA configuration for generating sequences of individual or collective sounds. Therefore, the ability of playing an instrument while composing is similar to writing a set of sequential instructions for caring out the musical composition.  
  
This paper has been divided into four sections. The first section is a literature review of ideas and studies associated with CAs and music. The second section is the Materials and Methods used for generating our CA and describing the data analysis. The third section shall show findings of our data analysis. The four section shall display the discussion and conclude with final remarks.

\section{Literature review}\label{LitRev}
This study is a theoretical approach for exploring alternative ways of generating music based on CAs. Compared with studies that translate their results of applying simple rules of computer programs to the music, we aim to define simple rules based on music theory for understanding and replicating aspects of human creativity. That is, writing simple programs based on the musical perspective complements the scientific approach for unveiling fundamental processes that occur in nature and in human capacity. Therefore, in this section, we shall briefly describe two proposed approaches for using CAs in music: the mathematical and the musical perspectives. These approaches are commonly related to the fields of social sciences and humanities, as well as the discussion between science and arts respectively.

As we pointed out in the last paragraph, the first approach is associated with studies of mathematical, statistical, and computational formulations. Under this perspective, the emphasis lies in translating complex patterns generated by mixing those formulations into a particular system of music. Because of the nature of its objective approach and its logical reasoning, the purpose of using this perspective is to create music from simple programs based on scientific rules and laws generally accepted by scientists. As we mentioned above, the work of Xenakis~\cite{Xenakis1992} and Wolfram~\cite{Wolfram2002} are the most representative of this approach. Following the ideas of these studies, we can identify the website of WolframTones~\cite{WolframTones2023} as one of the most complete option for exploring the application of computer programs and music theory in musical composition. In addition, there are other studies possibly related to this approach. The work of Miranda and Biles~\cite{MirandaandBiles2007} and Miranda~\cite{Miranda2021,Miranda2022} presented a vast and varied collection of documents related to computational modeling and music. In particular, the study of Miranda~\cite{Miranda2007} captured our attention because it describes the bases of using CAs in modeling music, and shows details of the CA sound Synthesis~\cite{Chareyron1990, Miranda1995, Roads1991}. Furthermore, the book edited by Miranda~\cite{Miranda2022} presented the field of quantum computing and its relationship with music. However, this first approach even today resembles an experimental method because it does not explicitly describe the possible ways of the creativity process. The approach does not consider the creativity as a scientific formulation for generating music due to its subjectivity. Consequently, studies associated with musical creativity have been little explored in the computer-aided process for recreating aspects of the human thinking.   

To overcome this limitation, the musical approach has considered the creativity as a fundamental part of conceiving a piece of music. Then, this approach is more appropriate for understanding and replicating aspects of the musical creativity because it concerns engaging human capacities with the technological developments in music. By using this perspective, we can generate computational models and analyze audio signals based on a particular system of music and the current music theory for identifying general principles associated with the musical creativity. For example, in the aspect of computer systems used in musical activities, the work of Pachet~\cite{Pachet2006}, which is part of a book edited by Deliege and Wiggins~\cite{DeliegeandWiggins2006}, addressed the creativity from a subjective point of view and provided an overview of non-traditional aspects of composing and performing music. In particular, he mentioned the ``style modeling programs'' as parts of computer systems that assist humans for generating music. That is, computer softwares assist composers and performers for creating their own musical material and showing a musical identity. In addition, in the compilation presented by Collins~\cite{Collins2012}, the work of Wiggins~\cite{Wiggins2012}  suggested that statistical models can be considered as efficient tools for predicting musical patterns related to music theory. Moreover, the study of Cope~\cite{Cope2012} showed the relationship between playing board games and computer programs to compose music. He suggested that the application of ``rules, tactics and strategies (RTS)'' in both events can be similar for successfully transitioning from games to computer programs.

In the aspect of analyzing audio signals, the work of Davy~\cite{Davy2006}, which is part of the book edited by Klapuri and Davy~\cite{KlapuriandDavy2006}, presented an overview of current signal processing methods for representing and analyzing musical data. In the same way, the work of Lugo and Alatriste-Contreras~\cite{LugoandAlatristeContreras2022} studied the spectrum of audio signals of the collaborative work between outstanding musicians who played music composed by Marin Marais. They analyzed the audio works named ``Marin Marais: Pièces de viole des Cinq Livres''~\cite{SavallKoopmanSmithCoinGallet2010} and identified statistical patterns associated with those audio signals. They found that the best-fit statistical distributions related to those signals were skewed statistical distributions, and they suggested that these type of distributions are the most common statistical attributes in the nature of audio signals related to music. Therefore, this musical approach may consider two distinct methodological formulations designed to analyze and generate music. Both of them complement each other and move towards the objective of better understanding the music and its creativity process.

In the next section, we shall present materials that describe the musical data. In particular, we mention the musical system used in this study and the Python libraries for coding and analyzing data.

\section{Material and methods}
\subsection{Musical and programing sources}
\begin{figure}[h]
\centering
\includegraphics[width=0.9\textwidth]{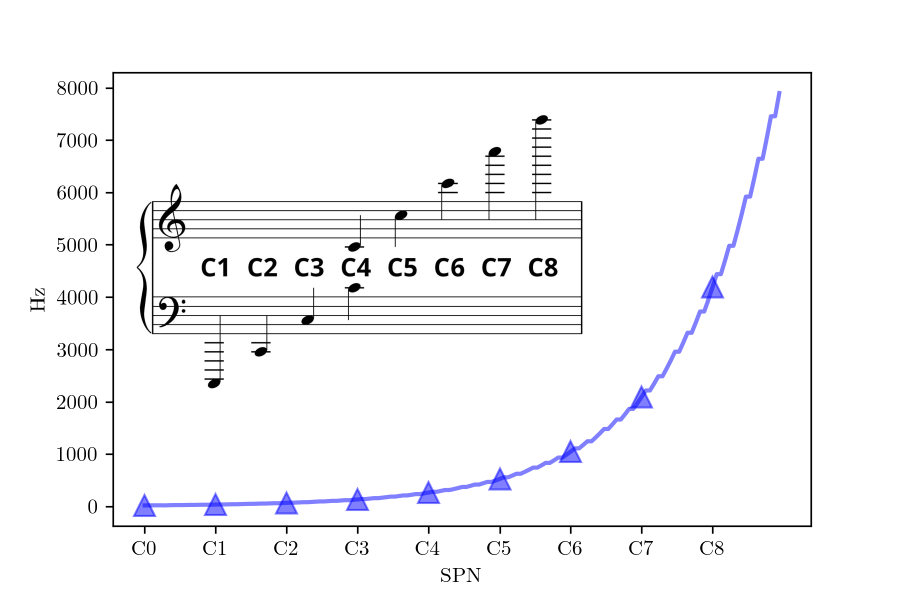}
\caption{Scientific pitch notation and frequencies based on the the twelve-tone equal temperament in which A4 = 440hz}\label{fig1}
\end{figure}

We used data from two sources. The first is the scientific pitch notation (SPN) that describes the sounds by tones (from A to G) following by a number related to an octave (Figure~\ref{fig1}). In particular, we used the Western system of the twelve-tone equal temperament in which A4 = 440hz (\href{https://pages.mtu.edu/~suits/notefreqs.html}{https://pages.mtu.edu/~suits/notefreqs.html}). Based on this data, we can elaborate the study by setting the initial conditions related to the state of cell values and the transitional rules in our CA.

The second source is related to Python libraries. In particular, we used the most common libraries such as Matplotlib (\href{https://matplotlib.org/}{https://matplotlib.org/}), Numpy (\href{https://numpy.org/}{https://numpy.org/}), and Scipy (\href{https://scipy.org/}{https://scipy.org/}). In addition, we used the Musicpy library (\href{https://musicpy.readthedocs.io/en/latest/}{https://musicpy.readthedocs.io/en/latest/}) because it shows a concise syntax for writing code and generating music. 

Therefore, based on those sources, we can generate our CA for exploring the effect of rules based on musical intervals in an array of random notes. Furthermore, the data and code for generating, analyzing, and replicating our CA is available as a supplementary material in the following public data repository: Open Science Framework (OSF), project \href{https://osf.io/vdqzu/?view_only=0e605bb1f32943d3bc6dc2ec5c3092f9}{Cellular automaton and music}. 

\subsection{Methods}\label{Meth}
Following the complex systems, we can generate and develop models using a particular musical system. In this respect, we can define the musical composition as the result of combining simple notes of the SPN in musical time and space for generating sequences of complex patterns that unified the musical material. Such combinations are related to singular mechanisms used by the human to associate musical components that are part of a such system of music and musical theory. Therefore, CAs may provide similar characteristics as traditional instruments in terms of intervals for composing and playing music. For example, an elementary CA or one-dimensional CA can be related to compose melodies based on melodic intervals---two notes sound one after another. To generate such a sequence of notes, we need to specify deterministic rules based on
the nearest neighbor notes for updating the current note. One of the most useful rule to set those neighbors is the use of intervals associated with a key tone. In addition, if we use a 2D CA, we can incorporate not only melodic but also harmonic interval. That is, we can include notes that sound at the same time. This addition can generate melodies and their harmonies. Thus, CAs in music composition can be seen as a natural extension for modeling music and replicating aspects of the musical creativity by identifying and translating the process of generating new ideas and their applications into particular rules. In this section we shall describe the fundamentals for generating and executing our CA proposal.

As we mentioned earlier, 2D CAs consist of a regular array in which each cell is associated with particular values called \verb+State+. The state of each cell shall be updated simultaneously over a closed boundary---a finite plane---based on simple rules that depend on its own state and its neighbor cells. We followed the formulation of the Game of Life for setting such rules due to they are simple and resemble real-life processes~\cite{Gardner1970}. Consequently, we defined the following main variables and updated rules:

\begin{enumerate}
	\item \verb+State+. It is a variable associated with the value of each cell in the array. The cell may show the value related to the SPN, for example, $A4$ or $F\#4$. This value shall be updated in discrete time steps according to the set of rules that depend on its neighborhood.
	\item \verb+Neighborhood+. It is associated with one of the two most common types of neighborhoods: Moore (Figure~\ref{fig2}). It is defined by the fourth orthogonal and diagonal adjacent cells. This neighborhood is used in a finite array, in a rectangular plane.
	
\begin{figure}[h]
\centering
\includegraphics[width=0.8\textwidth]{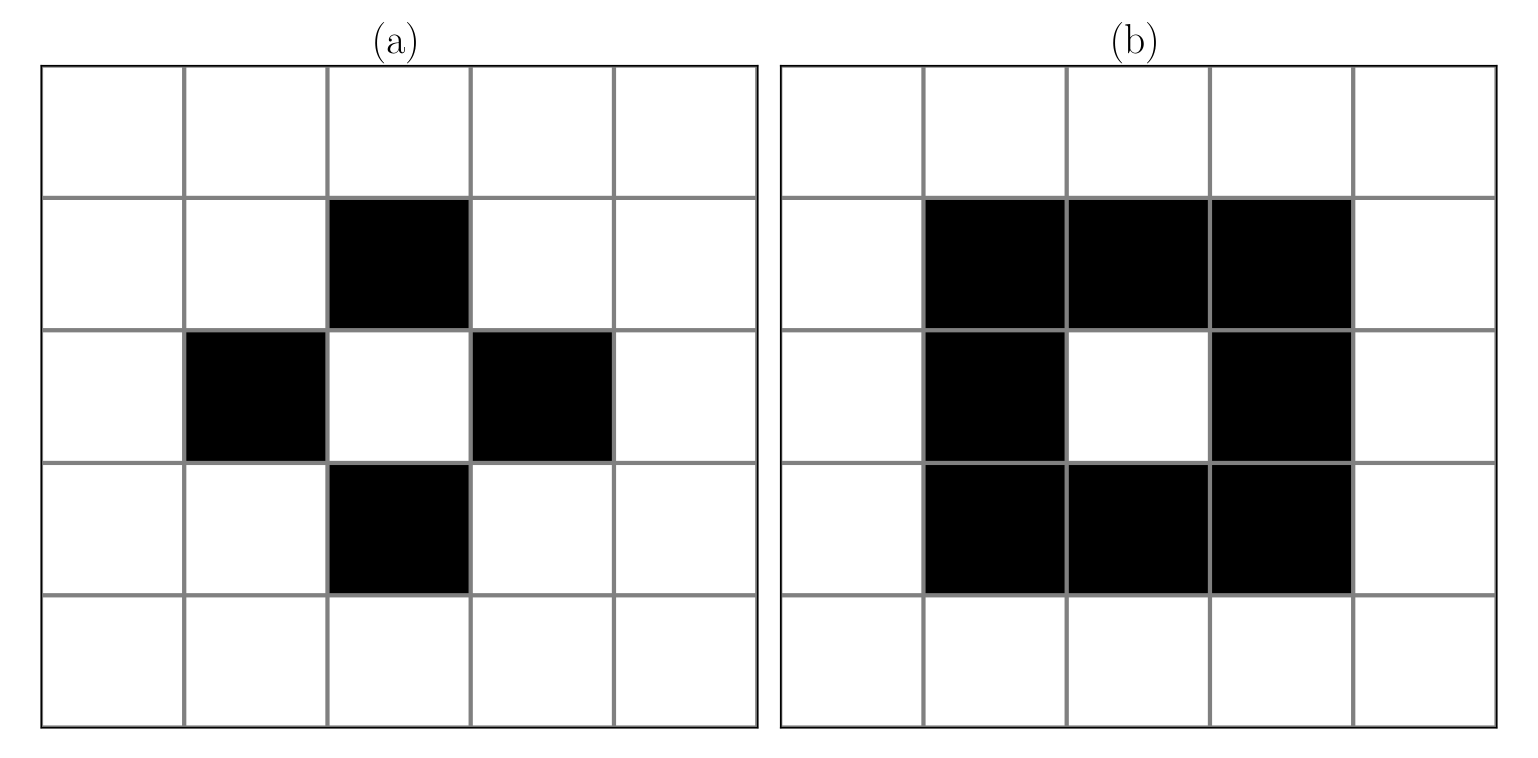}
\caption{Neighborhods. (a) Von Neuman and (b) Moore.}\label{fig2}
\end{figure}
	
	\item First and Second melodic intervals. These are two variables related to the key tone. The \verb+first set+ of melodic intervals is defined by the tuple of (root, 3rd, 5th, 7th) and the \verb+second set+ is defined by (2nd, 4th, 6th). The first interval is the common three notes in consecutive thirds (triads), plus the seventh note. The second set is related to those notes associated with the extended, suspended or added notes. For example, if we continue adding the similar interval of odd numbers after the 7th note, we extended the notes to 9th, 11th, and 13th. On the other hand, if we joint notes to the first intervals based on even numbers, we suspended the 2nd and 4th or added the 6th.
	\item \verb+Rules+. There are four musical rules that determine the update of each cell (Algorithm~\ref{rules1}). These rules update the whole array simultaneously:
	\begin{enumerate}
		\item STAY FLAT. It consists of repeating the same note. Each cell within the first or second sets of melodic intervals and with three or more neighboring cells within the first or second sets of melodic intervals updates its state with the current state.
		\item LEAP EMOTION. It refers to notes that go up and down on a scale. Each cell within the first or second sets of melodic intervals and with two or one neighboring cells within the first or second sets of melodic intervals updates its state by randomly selecting one value of the first or second set of melodic intervals.
		\item EXTEND, SUSPEND OR ADD. It consists of using extended notes (9th, 11th, 13th), or suspend/added notes (2nd, 4th, 6th). Each cell within the first or second sets of melodic intervals and with zero neighboring cells within the first or second sets of melodic intervals updates its state by randomly selecting one value of the first or second set of melodic intervals.
		\item SEARCH FOR SCALE TONE. It refers to notes randomly selected based on the neighboring cells. Each cell outside the first or second sets of melodic intervals updates its state by randomly selecting one value related to its neighboring cells.
	\end{enumerate}
	
\end{enumerate}

\begin{algorithm}
\caption{Musical rule}\label{rules1}
\begin{algorithmic}[1]

\If{state in first set}
	\If{neighbors $>= 3$ in first set}\\
		\hspace{1cm}state $=$ current state 
		\Comment{STAY FLAT}
	\ElsIf{neighbors $== 1$or $== 2$ state cells in first set}\\
			\hspace{1cm}state $=$ randomly selecting one value of first set
			\Comment{LEAP EMOTION}
	\ElsIf{neighbors $== 0$ state cells in first set}\\
		\hspace{1cm}state $=$ randomly selecting one value of second	set 
		\Comment{EXTENSIONS}
	\EndIf
\ElsIf{state in second set}
	\If{neighbors $>= 3$ in second set}\\
		\hspace{1cm}state $=$ current state 
		\Comment{STAY FLAT}
	\ElsIf{neighbors $== 1$or $== 2$ state cells in first set}\\
			\hspace{1cm}state $=$ randomly selecting one value of second set
			\Comment{LEAP EMOTION}
	\ElsIf{neighbors $== 0$ state cells in first set}\\
		\hspace{1cm}state $=$ randomly selecting one value of first set 
		\Comment{SUSPENDED/ ADDED}
	\EndIf
\Else \\
	\hspace{0.5cm}state $=$ randomly selecting one state of neighbors
	\Comment{SEARCH SCALE TONE}
\EndIf
\end{algorithmic}
\end{algorithm}

Therefore, based on these variables and rules, we can generate and execute our CA. In the next section, we describe the initial conditions and the type of data analysis used for validating our model.

\subsection{Initial conditions and data analysis}\label{InitCondDataAna}
Designed for simplicity, we used an array with a dimension (50, 50), $2500$ total number of cells. Each cell has a note related to the tone name in the SPN, from C0 to B8. The selection of each note is based on a random choice based on a uniform distribution that generates a random sample from the range of all possible notes in the SPN. To assure reproducibility, we used a random seed based on a random generator in the Numpy library (\href{https://numpy.org/doc/stable/reference/random/generator.html}{https://numpy.org/doc/stable/reference/random/\\generator.html}).
We updated the whole array 30 times to describe the dynamics of each simulation. It is important to mention that after this number of iterations, the model converges into a fixed pattern---from a disordered initial state to self-organized, order structures.

To validate our model, we present three cases associated with different types of rules. One of these cases is related to the rules proposed in the Algorithm~\ref{rules1}. The other two are associated with random and deterministic rules (Algorithm~\ref{rules2} and \ref{rules3}). The random rule presents a process in which the transition from the current cell value to the next one is related to a random selection of values associated with the SPN. This selection does not consider the current state of the cell and its neighboring values for updating the current value of a cell. On the other hand, the deterministic rule uses a transition that considers the neighboring values of a cell to update the current value of it. Such neighbors are related to a particular key tone defined ex-ante. Based on these neighboring values, the current cell updates its state by a particular note, $A4$. It is a deterministic rule because no randomness is involved in updating the future states~\cite{KerrGoethel2014}. Therefore, these cases intend to show a conceptual model validation in which our musical rule exemplify a case localized between probabilistic and deterministic frameworks~\cite{Sargent1984}. 

\begin{algorithm}
\caption{Random rule}\label{rules2}
\begin{algorithmic}[1]

	\Require $state$ = current value of a cell
	\State $state$ $=$ randomly selected one value associated with the SPN
	
\end{algorithmic}
\end{algorithm}

\begin{algorithm}
\caption{Deterministic rule}\label{rules3}
\begin{algorithmic}[1]

	\Require $state$ = current value of a cell
	\If{neighbors $>= 5$ in the key tone}
		\hspace{1cm} \State $state$ $=$ $A4$	
	\EndIf
	
\end{algorithmic}
\end{algorithm}

Based on these cases for validating the CA model, we use an explorative data analysis for identifying similarity or dissimilarity patterns. In particular, we use the last realization of the updated array for comparing each case along with others. Based on this data, we apply the Spearman correlation coefficient. Next, we identify the best fit statistical distribution based on the Kolmogorov--Smirnov (KS) test for goodness of fit that best describes the data associated with those updated arrays~\cite{Massey1951}. We use a set of continuous statistical distributions for identifying the best fit of our data~\cite{LugoAlatristeContreras2019} (for more information about the set of those distributions, see the supporting information, OSF, project \href{https://osf.io/vdqzu/?view_only=0e605bb1f32943d3bc6dc2ec5c3092f9}{Cellular automaton and music}). Finally, we present the estimated parameters of the statistical distribution related to the best fit analysis, as well as their statistical moments.

\section{Results}\label{Res}
As we mention earlier, this type of CA may replicate some aspects of the creativity because it showed how random arrays of notes were organized by using local rules based on the music theory. This ex-ante random condition resembles those common circumstances that musicians and non musicians in a traditional approach face during the process of musical composition. For example, the mental activity for generating new ideas and executing them is based on existed knowledge that the individual uses for creating music. Therefore, based on the Western system of the twelve-tone equal temperament and the music theory the individual can produce new music based on existing ideas and concepts. In this section, we shall present our findings in which the current musical system and their rules produce musical patterns that closely resemble the pleasing music. In particular, we shall show the results of the three cases mentioned in the last section associated with the different rules of transition: random, musical, and deterministic rules. 

We begin with the visual finding of the proportion of cells that are associated with notes in the key tone (Figure~\ref{fig2}). We have to remind that the initial configuration of the random array is the same in all three cases, and we used $Emin$ as our key tone. Then, Figure~\ref{fig2} shows each case in which the number of cells is related to the particular key tone in every iteration.

\begin{figure}[h]
\centering
\includegraphics[width=0.85\textwidth]{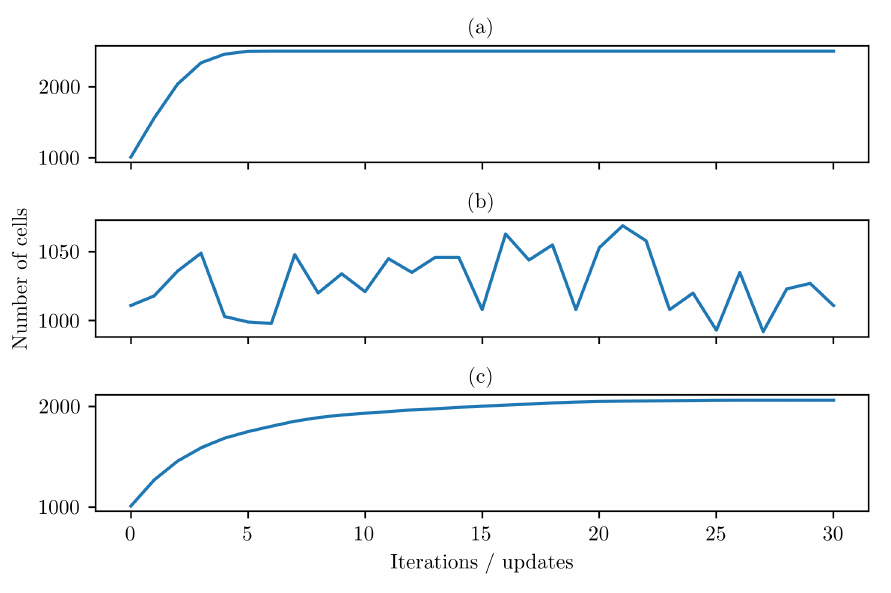}
\caption{Number of cells and iterations in each case. (a) Musical, (b) Random, and (c) Deterministic}\label{fig2}
\end{figure}

The three subfigures show different trends in the proportion of cells associated with notes in the key tone. 
Subfigure $(a)$ shows a sharp rise in the number of cells in the tonal key. We can see that after 5 iterations the number of cells in the key tone is almost complete, i.e., the array has been harmonized.
Subfigure $(b)$ displays the common behavior of random numbers generated by a uniform distribution in which the number of cells in the tonal key shows a marked rise and fall.
Subfigure $(c)$ displays a gradual rise in the number of cells associated with the tonal key. We can see that reached the 30 iteration the total number of cells in the key is closed to 80\% of the cells, approximately $2000$ cells.
Therefore, based on the transitional rules, the three cases showed different dynamics in the number of cells associated with the key tone.

Next, we present the results of the Spearman correlation (Table \ref{tab1}). This table provides the statistics for every combination of pair of cases. We can see in Table~\ref{tab1} that each pair of cases display small correlation coefficients---small in magnitude, close to zero. Therefore, the null hypothesis that two cases have no ordinal correlation is accepted. 

\begin{table}[h]
\caption{Spearman correlation between cases}\label{tab1}%
\centering
\begin{tabular}{|c|c|c|}
\hline
Pair of cases & Statistic  & p-value \\
\hline
Musical and Random    & 0.016002   & 0.423846  \\
Musical and Deterministic    & 0.038947   & 0.051516  \\
Random and Deterministic    & 0.019656  & 0.325886   \\
\hline
\end{tabular}
\footnotetext{The data per case is related to the last updated array.}
\end{table}

After, we identified the statistical distribution that best describes the data in each case (Table~\ref{tab2}). As we can see in the Table~\ref{tab2}, the three cases show the gamma distribution as the statistical distribution that best describes each data set. In addition, we can see the estimated parameters of the gamma distributions and their moments~(Table \ref{tab3} and \ref{tab4}).

\begin{table}[h]
\caption{Best fit analysis per case}\label{tab2}%
\centering
\begin{tabular}{|c|c|c|c|}
\hline
Case & Distribution  & KS test & p-value\\
\hline
Musical    & gamma   & 0.026060  & 0.065857  \\
Random    & gamma   & 0.027161  & 0.049090 \\
Deterministic    & gamma   & 0.026682  & 0.055872 \\
\hline
\end{tabular}
\footnotetext{The data per case is related to the last updated array. After using the KS test in our set of continuous statistical distributions, we found that the statistics of the gamma distribution were the most accurate. See the supplementary information regarding the KS test used in the set of continuous statistical distributions.}
\end{table}

\begin{table}[h]
\caption{Best fit distribution per case and its estimated parameters}\label{tab3}%
\centering
\begin{tabular}{|c|c|c|}
\hline
Case & Distribution  & Estimated parameters (a, loc, and scale) \\
\hline
Musical    & gamma   & (0.416808, 16.349999, 1824.393176)   \\
Random    & gamma   & (0.511168, 16.349999, 2152.674516)   \\
Deterministic    & gamma   & (0.440319, 16.349999, 1905.617941)  \\
\hline
\end{tabular}
\footnotetext{The data per case is related to the last updated array.}
\end{table}

Table~\ref{tab3} shows different values in the estimated shape parameters. However, these values are relatively close in magnitude. The estimated location parameters are identical in each case, meanwhile the estimated scale parameter varies. Therefore, the estimated parameters, in particular the shape parameter, describes highly skewed shapes.

\begin{table}[h]
\caption{Best fit distribution per case and its the median and the first four moments}\label{tab4}%
\centering
\begin{tabular}{|c|c|c|c|c|c|c|}
\hline
Case  & Distribution &median & mean & variance & skewness & kurtosis\\
\hline
Musical  & gamma &996.822396  & 776.772721 & 1387310.023786 & 3.097859& 14.395097\\
Random  & gamma &1492.648467 & 1116.729061 & 2368757.964797 & 2.797357 & 11.737816\\
Deterministic  & gamma &1110.519330 & 855.431465 & 1598968.694566 & 3.014018 & 13.626457\\
\hline
\end{tabular}
\footnotetext{The data per case is related to the last updated array.}
\end{table}

Finally, Table~\ref{tab4} displays the moments of the data related to the best fit analysis. In the variance measures, we can see that the rule of transition related to the random rule shows the greatest variability. On the other hand, the transition related to the musical rule shows the smallest variability. In terms of symmetry, the three cases show positive skewness in which the musical transition displays the greatest asymmetry, and the random rule shows the smallest asymmetry. In the same vein, the three cases show positive excess of kurtosis that indicates fatter tails. In particular, the case of the musical rule displays the higher value of kurtosis, and the random case shows a relative smaller kurtosis measure. Therefore, these findings suggested that the gamma distribution that best describes our data is associated with skewed shapes and fatter tails. 

In summary, these findings showed clear differences between the three cases. The visual and Spearman findings indicated that the three cases are significantly different from each other. Meanwhile, in the best fit analysis, we showed similarities in the statistical distribution that best describe the final actualization of data, but dissimilarities in the estimated parameters and the moments of such statistical distributions.  

\section{Discussion}\label{Disc}
Musical composition is the output of the musical creativity, which is the mental activity for generating music based on existed music systems and theories. Based on this idea and our findings, we could identified two similarities and one dissimilarity between the traditional and our computational approaches for composing music.

Similarities are related to our research questions in which the first question sought to determine the relevance of the transitional rules based on music theory and a particular key tone for coordinating large-scale sets of random notes. CAs can coordinate different sets of random notes if they follow particular music intervals and a particular key tone. Compared with the cases of random and deterministic rules, the data of our musical rule showed the smallest variability and the greatest asymmetry and kurtosis. These findings suggest that the sounds that we perceive as music are related to highly skewed statistical distributions. In this case, the gamma distribution associated with the musical rule shows the natural characteristic of audio signals related to music.

The other similarity is associated with the second research question which sought to address the relationship between the transitional rules based on music intervals in a CA and its closeness with the traditional approach for composing music. Such transitional rules can replicate one aspect of the creativity because in both approaches there is a human who determines different strategies for creating music by using the musical notation, played instruments, or writing computer code. Based on our findings related to the three cases, in particular the musical rule, we could see that by following particular music intervals, which are related to a particular music system and theory, we can generate outputs closely related to musical sounds. In the other cases, the output is closely related to noise sounds. 

The dissimilarity that we could identify is the use of computers, in particular the CA modeling, as an instrument for producing music. Compared with the traditional approach for composing music, the CA modeling approach uses the computer as an instrument for producing musical sounds. Therefore, the programer who generates the CA may use all the musical resources at hand, such as the music theory, and the most successful strategies for translating the musical intuition and experience to simple algorithms. 

Therefore, these findings are significant in at least two major respects. The first is related to the application of the CA in music. Based on music theory, the CAs may produce pleasant and organized sounds closely related to music. In addition, we can analyze these sounds as an audio signal for identifying common patterns that define the nature of music, such as the statistical distribution that best describes the audio signal and its statistical moments. 
The second is related to the creativity process that produces the musical composition. In this respect, our CA suggests that the musical creativity required to compose music can be expressed in simple lines of code. Therefore, no matter what musical approach someone uses for composing music, the output will be similar.

Future work is needed to evaluate our musical approach in other types of computational models. In particular, the use of complex networks for exploring different ways of coordinating a set of random notes and types of musical creativity behind it.

\subsection{Conclusion}
In this investigation, the aim was to assess the applicability of a 2D CA based on music intervals for coordinating a random array of notes. The use of transitional rules associated with music theory and the possible coordination of the initial array of random notes suggest that this type of computational model can replicate some aspects of the creativity used in the musical composition. Therefore, CAs based on music theory have the potential to complement scientific and humanities studies for understanding the musical composition and replicating the music creativity. 

\section*{Supplementary information}
Supplementary information is available in the [Open Science Framework] repository, [\href{https://osf.io/vdqzu/?view_only=0e605bb1f32943d3bc6dc2ec5c3092f9}{https://osf.io/vdqzu/?view\_only=\\0e605bb1f32943d3bc6dc2ec5c3092f9}].

\section*{Declarations}

\begin{itemize}
\item Funding. Not applicable.
\item Conflict of interest/Competing interests. The authors have no competing interests to declare.
\item Ethics approval. Not applicable.
\item Consent to participate. Not applicable.
\item Consent for publication. Not applicable.
\item Availability of data and materials. The datasets used and generated and/or analyzed during the current study are available in the [Open Science Framework] repository, [\href{https://osf.io/vdqzu/?view_only=0e605bb1f32943d3bc6dc2ec5c3092f9}{https://osf.io/vdqzu/?view\_only=\\0e605bb1f32943d3bc6dc2ec5c3092f9}].
\item Code availability. The underlying code [and training/validation datasets] for this study is available in [Open Science Framework] and can be accessed via this link [\href{https://osf.io/vdqzu/?view_only=0e605bb1f32943d3bc6dc2ec5c3092f9}{https://osf.io/vdqzu/?view\_only=\\0e605bb1f32943d3bc6dc2ec5c3092f9}]. 
\item Authors' contributions. IL contributes with the following parts: conceptualization, formal analysis, investigation, methodology, validation, visualization, writing---original draft, Writing---review \& editing.
MGA-C contributes with the following parts: conceptualization, formal analysis, methodology, validation, writing---original draft, Writing---review \& editing.
All authors read and approved the final manuscript.
\end{itemize}



\end{document}